\documentclass[letterpaper,pre,twocolumn,floatfix,superscriptaddress]{revtex4}  
\usepackage[normalem]{ulem}  
\usepackage{amsmath}
\usepackage[mathcal]{euscript}
\usepackage{mathrsfs}
\usepackage{float}
\usepackage[usenames,pdftex]{color}
\usepackage[pdftex]{color,graphicx}     
\usepackage[pdftex,bookmarks=false]{hyperref}           

\newcommand{\x}{\mathbf{x}}

\begin{document}

\title{Constructing explicit magnetic analogies for the dynamics of glass forming liquids}
\date{\today}
\author{Jacob D. Stevenson}
\affiliation{Department of Physics and Department of Chemistry and Biochemistry,
University of California, San Diego, La Jolla, CA 92093}
\author{Aleksandra M. Walczak}
\affiliation{Princeton Center for Theoretical Physics, Princeton University, Princeton, NJ 08544}
\author{Randall W. Hall}
\affiliation{Department of Chemistry, Louisiana State University, Baton Rouge, Louisiana 70803-1804}
\author{Peter G. Wolynes}
\affiliation{Department of Physics and Department of Chemistry and Biochemistry,
University of California, San Diego, La Jolla, CA 92093}
\affiliation{e-mail: pwolynes@ucsd.edu}

\begin{abstract} 
  By defining a spatially varying replica overlap parameter for a supercooled
  liquid referenced to an ensemble of fiducial liquid state configurations we
  explicitly construct a constrained replica free energy functional that maps
  directly onto an Ising Hamiltonian with both random fields and random
  interactions whose statistics depend on liquid structure.  Renormalization
  group results for random magnets when combined with these statistics for the
  Lennard-Jones glass suggest that discontinuous replica symmetry breaking
  would occur if a liquid with short range interactions could be equilibrated
  at a sufficiently low temperature where its mean field configurational
  entropy would vanish, even though the system strictly retains a finite
  configurational entropy.
\end{abstract}

\maketitle

\bibliographystyle{/people/jake/latex_files/naturemagurl}

The commonalities and contrasts between the glassy behavior of supercooled
liquids and quenched disordered magnetic systems have been long studied.  Both
exhibit a diversity of long lived states with no apparent structural long range
order.  These states challenge the paradigms of standard many body physics.
Starting in the 1980's, the disordered magnetic systems came under control, but
only in mean field, through the concept of broken ergodicity using replica
methods which allowed averaging over quenched disorder.  Despite the absence of
quenched disorder there is an analogy between the first order breaking of
ergodicity predicted by mode coupling theory for structural glasses and the
phase transition predicted for spin glasses that lack up-down symmetry (Potts
spin glasses)\cite{kirkpatrick.1987a,kirkpatrick.1987c}.  In mean field, the
asymmetric spin glasses exhibit a Kauzmann entropy
crisis\cite{kirkpatrick.1987b} like that for supercooled liquids. 

The analogy between glass forming liquids and mean field spin glasses is
however incomplete.  The observed dynamics of supercooled liquids fits the mean
field mode coupling paradigm only modestly well\cite{das.2004}.  The main
problem with the mean field theory is its neglect of important activated
motions that allow a supercooled liquid to re-configure locally.  The most
convincing evidence for this is the near Arrhenius dynamics with large
activation energy seen in structural glasses in the aging
regime\cite{lubchenko.2007}.

Activated motions have infinite barriers in mean field.  Describing activated
motions requires accounting for the finite interaction range.  Accompanying a
paucity of exact results\cite{young.1997}, even the empirical situation remains
controversial for finite range random magnetic
systems\cite{marinari.2000,parisi.2007}.  For the random field Ising magnet,
arguments based on renormalization group theory and droplet arguments do work
well\cite{huse.1987,mcmillan.1985,villain.1985},  but controversy remains as to
the extent Ising spin glasses partake of mean field versus droplet
features\cite{huse.1991}.  Nevertheless, useful analogies between glass forming
liquids and disordered magnets have been drawn showing how droplet arguments
give the Vogel-Fulcher law\cite{kirkpatrick.1989}.  Similar arguments have also
been adduced using replicas\cite{kirkpatrick.1987b,dzero.2005,franz.1995}.  

By applying density functional theory to determine the parameters in these
droplet arguments\cite{xia.2000} one predicts a large number of confirmed
quantitative results for liquids\cite{lubchenko.2007}.  The resulting random
first order transition (RFOT) theory bears some resemblance to the nucleation
picture of ordinary first order phase transitions.  The theory's core is an
analogy to the random field Ising magnet in a field. The average field in the
magnet is related to the configurational entropy density of the
liquid\cite{kirkpatrick.1989,xia.2000}.  Tarzia and Moore\cite{tarzia.2007}
suggest that supercooled liquids are related to Ising spin glasses in a field.
When an average field is present, corresponding to a finite configurational
entropy in the liquid, both the random field magnet and the spin glass have the
same symmetry: neither model is expected to have a phase transition although
this is still somewhat controversial\cite{temesvari.2008}.  While both magnetic
analogies agree on this point, the activated dynamics in the two different
analogical magnets differ because the interface energies in spin glass droplets
have weaker scaling than the random field ferromagnet. To quantify this
distinction in this paper we construct explicitly the analogy between a
structural glass forming liquid and the corresponding short range disordered
ferromagnet.  

Replica methods along with liquid state theory allow an explicit mapping of the
free energy landscape of a glass forming liquid onto a disordered Ising magnet.
Using this explicit construction we may eschew droplet arguments entirely.
Simulating the analogous magnet shows the mapping reproduces, however, the main
features of the RFOT analysis.  The mean activation barrier and its
fluctuations which give rise to non-exponential relaxation can be computed
using importance sampling methods for the analog.  Approximate arguments can
place the analog system onto a phase diagram previously deduced for disordered
magnets using renormalization group (RG).  This construction suggests that
under thermodynamic conditions when the mean field estimate for the
configurational entropy would vanish, the liquid would still undergo a phase
transition having one step replica symmetry breaking, despite the exact
configurational entropy remaining non-zero\cite{eastwood.2002,stillinger.2001}.

\section{Theory}

The analogy yields a description of a structural glass in terms of discrete
spin-like variables tied to the liquid structure equilibrated at one time. A
structural glass is statistically homogeneous but nonuniform with a density
$\rho (\x )$ that is not translationally invariant.  Liquid state theory
provides a free energy as a functional of such a density $\mathcal{F}[\rho
(\x)]$. While the complete equilibrium free energy $F=-k_BT \ln Z$ assumes all
phase space can be sampled, close to the glass transition there is trapping in
locally metastable states which manifests itself as extensively many local
minima described by a configurational entropy density $S_c=\ln \it{N}_{ms}$.
This non-ergodic behavior can be captured by a construction due to
Monasson\cite{monasson.1995} in which an external random constraining pinning
field couples $m$ replicas of the system through an attractive potential.  Each
replica's density field is $\rho^k(\x)$.  The free energy of the $m$ replica
system is

\begin{equation}
  \begin{split}
    &F (m, \beta)=\lim_{g\rightarrow 0+} \Bigg[  -\frac{1}{\beta m} \ln \int
    \prod_{k=1}^m  \mathcal{D} \rho^k(\x ) \exp \Bigg\{\\
    & -\beta \sum_k
    \mathcal{F}[\rho^k(\x )]  -\frac{g}{2m} \sum_{k,j, k<j}
    \int d \x  [\rho^k(\x )-\rho^j(\x )]^2\Bigg\}  \Bigg]
  \end{split}
  \label{replicaFE}
\end{equation}

The typical free energy of a metastable frozen state, $\tilde{F} =
\frac{\partial m F(m, \beta)}{\partial m}\big|_{m=1}$, differs from the
complete equilibrium free energy by an amount \mbox{$\delta F = \tilde{ F} - F
= \frac{\partial F(m,\beta)}{\partial m}\big|_{m=1} = TS_c$ }.  In contrast to
Monasson's thoroughly field theoretic formulation, we separate the $m$ replicas
into one fiducial probe copy of particles interacting through the Hamiltonian
$H(\{\x_i^f\}) = \sum_{i<j} u(\x_i^f - \x_j^f)$ and $m-1$ others described by
density fields.  Here $u$ is the microscopic inter-particle potential. The
pinning field on the other replicas is $\rho^f(\x )=\sum_i \delta(\x-\x_i^f)$.
We can write:

\begin{widetext}
\begin{equation}
  \begin{split}
    F(m,\beta) =& -\lim_{g\to0} \frac{1}{\beta m} \ln  \int \mathcal{D}
    \{\x_i^f\} e^{-\beta H\left(\{\x_i^f\}\right)} \int \prod_{k=1}^{m-1}
    \mathcal{D} \rho^k(\x ) \exp \Bigg\{-\beta \sum_{k=1}^{m-1}
    \mathcal{F}[\rho^k(\x )] \\ 
    &  -\frac{g}{2m} \sum_{k=1}^{m-1} \int d \x [\rho^k(\x )-\rho^f(\x )]^2
    -\frac{g}{2m} \sum^{m-1}_{k=1}\sum^{m-1}_{j=1, k<j} \int d \x [\rho^k(\x
    )-\rho^j(\x)]^2 \Bigg\}
  \end{split}
  \label{eqn:2}
\end{equation}
\end{widetext}

In calculating the free energy with respect to the probe replica we ignore
peripheral interactions not involving the probe, thus decoupling the partition
function

\begin{equation}
  F(m,\beta)= -\lim_{g\to0} \frac{1}{\beta m} \ln \int \mathcal{D} \{\x_i^f\} e^{-\beta
  \mathcal{H}\left(\{\x_i^f\}\right)} [Z_q]^{m-1}
\end{equation}

\noindent such that each replica is constrained by a potential to the vicinity of
the fiducial copy:

\begin{equation}
  \begin{split}
  Z_q= \int  &\mathcal{D} \rho(\x )   \exp \bigg\{ -\beta \mathcal{F}[\rho(\x )] \\
  &-\frac{g}{2m} \int d \x [\rho(\x )-\rho^f(\x )]^2\bigg\}
  \label{eqn:zq}
  \end{split}
\end{equation}

For a glassy system, when $TS_c =  \frac{\partial F(m, \beta)}{\partial
m}|_{m=1}$ is finite, in the limit $m \rightarrow 1$, the replicated free
energy is dominated by a saddle point corresponding to the spontaneous ordering
of replicas in phase space.  The saddle point solution to the free energy is
found by minimizing the exponential's argument.

\begin{equation}
  \frac{\delta}{\delta \rho(\x )} \left[ \beta
  \mathcal{F}[\rho(\x )] + \frac{g}{2m} \int d \x 
  [\rho(\x )- \rho^f (\x ) ]^2
  \right]_{\rho=\tilde{\rho}}=0
  \label{minimize}
\end{equation}

\noindent The free energy functional can be written, \`{a} la density
functional theory, as the sum of an entropic cost to localize the density and
an interaction term.

\begin{equation}
  \beta \mathcal{F}[\rho(\x)] \approx \int d\x \rho(\x) \ln \rho(\x)  + \beta
  F_{int}[\rho(\x)].
  \label{}
\end{equation}

\begin{figure}[tpb]
  \begin{center}
    $\begin{array}{c}
      \multicolumn{1}{l}{\mbox{\bf (a)}} \\ [-0.53cm]
      \includegraphics[width=0.48\textwidth]{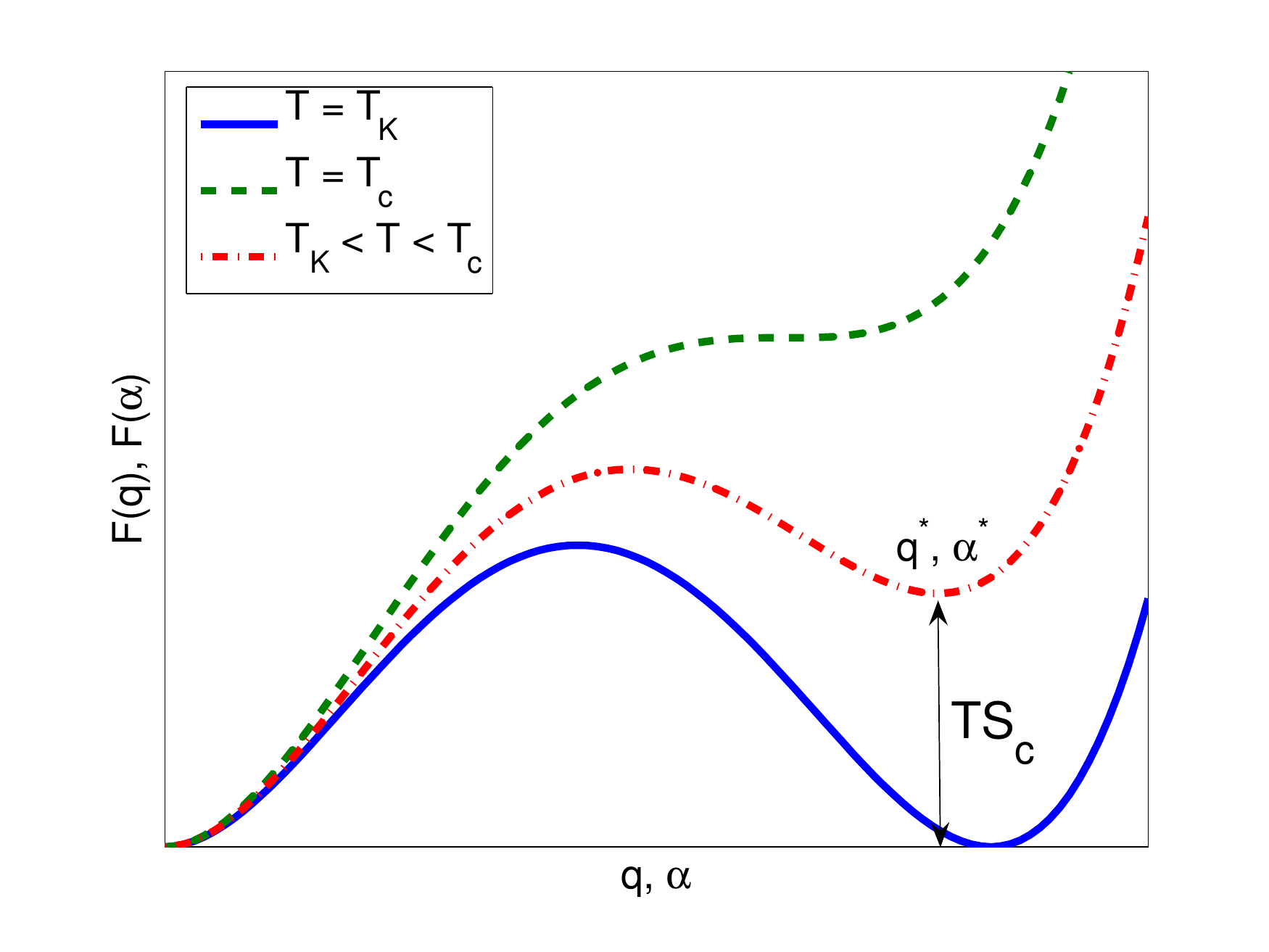} \\
      \multicolumn{1}{l}{\mbox{\bf (b)}} \\ [-0.53cm]
      \includegraphics[width=0.48\textwidth]{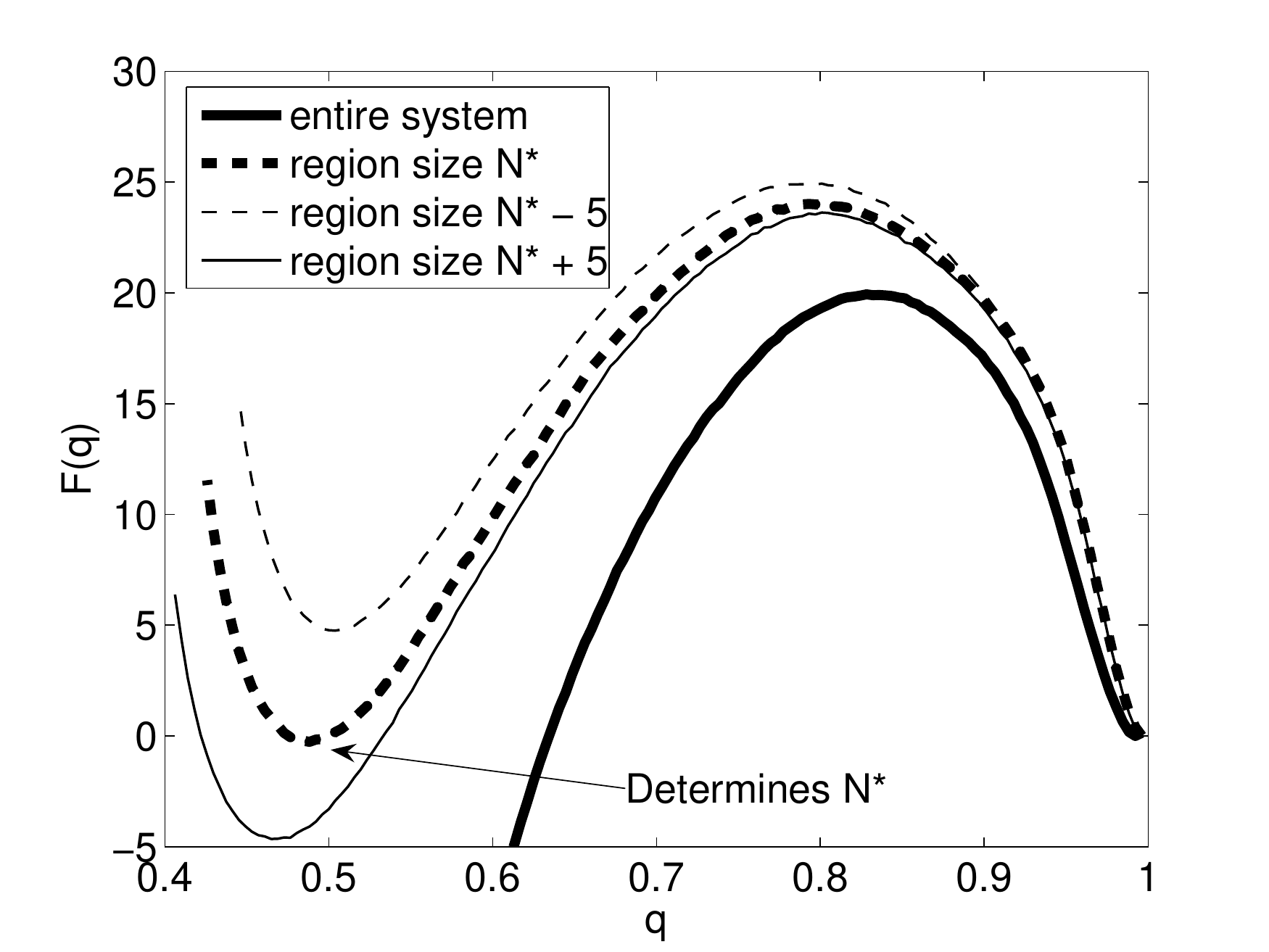}\\
    \end{array}$
  \end{center}
  \caption{(a) Schematic mean field free energy profiles for supercooled
  liquids at the dynamical crossover temperature (dashed line), the Kauzmann
  temperature (solid line), and an intermediate temperature (dot-dashed line).
  In mean field the particle localization, $\alpha$, and the structural
  overlap, $q$, are equivalent reaction coordinates.  The secondary free energy
  minimum at $T_K < T < T_c$ demonstrates the existence of metastable
  structural states in supercooled liquids. (b) Free energy profiles calculated
  for the finite range Ising magnet analogous to the LJ liquid.  The minimum
  size needed to escape the free energy minimum and thus reconfigure the liquid
  at $s_c = 1.10$ is $N^{*} = 148$ particles.}
  \label{fig:Fqschematic}
\end{figure}

\noindent While the free energy functional $\mathcal{F}[\rho]$ is globally
minimized by a uniform equilibrium solution with mean density $\rho_0$,
$\mathcal{F}[\rho]$ also has local minima corresponding to frozen aperiodic
densities.  The g coupling, even as g vanishes, picks out one particular
minimum around the structural state $\rho^f$.  Thus $\mathcal{F}[\rho]$ can be
analyzed in terms of the similarity, or overlap \mbox{$q = \int d\x (\rho(\x )
- \rho_0)(\rho^f(\x ) - \rho_0)$}, between $\rho(\x )$ and $\rho^f(\x )=\sum_i
\delta(\x-\x_i^f)$.  A schematic of the free energy as a function of the order
parameter $q$ as computed in references \cite{singh.1985} and \cite{dzero.2005}
is shown in figure \ref{fig:Fqschematic}a.  The free energy difference between
the large overlap solution and the small overlap solution is the excess free
energy of the frozen glass over the equilibrium free energy and is determined
by $TS_c$.

The large overlap state is well approximated by a density distribution of a sum
of Gaussians centered around the particle locations of the fiducial (probe)
copy \mbox{$\rho(\x ) = \sum_i \rho_i(x) =  \sum_i \left( \frac{\alpha_i}{\pi}
\right)^{3/2} e^{-\alpha_i (\x -\x_i^f)^2}$}.  The localization parameters, $\{
\alpha_i \}$, determine the local overlap.  Near the large overlap minimum,
\mbox{$q \big(\alpha_i \gg \rho_0^{2/3} \big) = \sum_i \left( (\alpha_i /
\pi)^{3/2} - \rho_0\right)$}.  In the opposite limit, near the global free
energy minimum, the density ansatz reduces to the mean density and $q( \{ \alpha_i \} \to 0) = 0$.

For large values of $\{ \alpha_i \}$ the particles are localized very near the
fiducial locations $\{\x_{i}^f\}$ and $F$ can be evaluated by using the
independent oscillator approximation\cite{fixman.1969,stoessel.1984} which
decouples the particles at the individual site level.  Within this
approximation $F_{int}$ can be expressed as a sum of effective potentials
between the interacting density clouds, \mbox{$\beta V_{eff} \left( | \x^f_i -
\x^f_j |; \alpha_j \right) \equiv - \ln \int d\x_j \rho_j(\x_j) e^{-\frac{1}{2}
\beta u(\x_i^f - \x_j)} $}.

\begin{equation}
  \begin{split}
  \beta F_{glass}  \Big( \{ \x_i^f \}, & \{ \alpha_i \}
  \Big) = \sum_i \frac{3}{2} \ln \frac{\alpha_i \Lambda^2}{\pi e} \\
  &+ \sum_{ij} \beta
  V_{eff} \left( | \x^f_i -  \x^f_j |; \alpha_j \right) .
  \label{eqn:Falpha1}
  \end{split}
\end{equation}

The localization parameters corresponding to the large overlap solution, $\{
\alpha_i^{\uparrow} \}$, can be found by applying a self consistency
condition\cite{fixman.1969,stoessel.1984}. The existence of the free energy
minimum at large overlap reflects the cage effect where the motion of a
particle is restricted by its neighbors.  To first order one may compute the
potential \mbox{$\beta V_e \left( \x_i - \x_i^f, \{\alpha_j\} \right) \equiv
\sum_j \beta V_{eff} \left( | \x_i -  \x^f_j |; \alpha_j \right) $} and expand
around small displacements of particle $i$, $w_i = | \x_i - \x_i^f |$.

\begin{equation}
  \beta V_{e}\left(w_i,\{\alpha_j^{\uparrow}\}\right)  \approx \left. \beta
  V_{e}\right|_{w_i = 0} + 
  w_i^2 \frac{1}{6} \nabla^2 \beta \left.
  V_{e}\right|_{w_i = 0}
  \label{eqn:v_expansion}
\end{equation}

\begin{equation}
  \beta V_{e}\left(w_i,\{\alpha_j^{\uparrow}\}\right)  \approx \beta \left.
  V_{e}\right|_{w_i = 0} + 
  \alpha^{\uparrow}_i w_i^2
  \label{eqn:alpha_def}
\end{equation}

\noindent The linear term gives no contribution because the fiducial replica
becomes centered on a stationary location with all forces canceling.  The
localization parameter of particle $i$ thus can be computed from the curvature
of the effective potential, \mbox{$\alpha_i^{\uparrow} = \frac{1}{6} \nabla^2
\beta V_e \left( | \x_i - \x_i^f | = 0,\{\alpha_j^{\uparrow}\}\right)$}, giving
a self-consistent solution for
$\alpha_i^{\uparrow}$\cite{mezard.1999,stoessel.1984}.

Near the uniformly low overlap state the interaction free energy $F_{int}$
follows from the equilibrium liquid equation of state, $Z_{EoS}(\eta)$, where
$\eta$ is the packing fraction\cite{hall.2008}:

\begin{equation}
  F_{liq}^{\downarrow} = N \ln \rho_0 \Lambda^3 - N + N \int_0^\eta (Z_{EoS} -
  1) \frac{d \eta'}{\eta'}.
\end{equation}

To characterize the reconfiguration events and develop the magnetic analogy one
must also examine non-uniform solutions. At the interface between the two
solutions there must be some energetic penalty due to the patching together of
distinct configurational states.  At an interface, one particle's replica is in
the large overlap state while it's neighbor has small overlap, so the pair
interaction becomes {$\beta V_2^{eff} \left( | \x^f_i -  \x^f_j |;
\alpha_i^{\downarrow}, \alpha_j^{\uparrow} \right) = - \ln \int d\x_i d\x_j
\rho_i^{\downarrow}(\x_i) \rho_j^{\uparrow}(\x_j) e^{-\frac{1}{2} \beta u(\x_i
- \x_j)} $}.  The small overlap parameters $\{ \alpha_i^{\downarrow} \}$
determine $\rho_i^{\downarrow}$ and are obtained in the self-consistent phonon
theory by matching the entropy of the low overlap state calculated within the
Gaussian density ansatz with the entropic term of the equilibrated liquid: 

\begin{equation}
  \sum_i \frac{3}{2} \ln \frac{\alpha_i^{\downarrow} \Lambda^2}{\pi e} = N \ln
  \rho_0 \Lambda^3.
  \label{eqn:low_alpha}
\end{equation}

For any combination of the discrete values of $\{ \alpha_i^{\downarrow} \}$ and
$\{ \alpha_i^{\uparrow} \}$, the free energy of the supercooled liquid is
equivalent to a pairwise interacting model with spins located at the fiducial
locations, $\{ \x_i^f \}$. 

\begin{equation}
  \beta H = - \sum_i h_i (1 - s_i) +
  \sum_{i<j} J_{ij} \left[ s_i (1 - s_j) + s_j (1 - s_i) \right],
  \label{eqn:ising_hamiltonian}
\end{equation}

\noindent where the spin, $s_i = 1$, corresponds with large overlap and $s_i =
0$ a small overlap site. The average field is found from the bulk free energy
difference between the states, \mbox{$\sum_i h_i = \beta F_{glass} - \beta
F_{liq} = Ns_c / k_B$}, with a heterogeneous local configurational entropy
resulting from the alpha variations.

\begin{equation}
  h_i = \frac{3}{2} \ln \frac{\alpha_i^{\uparrow}}{\pi} +
  \beta \sum_{j} V_{eff}(|\x_i^f - \x_j^f|; \alpha_j^{\uparrow})
  - \frac{1}{N} F_{liq}
\end{equation}

\noindent The interactions defined through the effective potential
give the surface energies of droplets within the RFOT
picture and are explicitly

\begin{equation}
  J_{ij} = V_2^{eff} \left( | \x^f_i -  \x^f_j |;
  \alpha_i^{\downarrow}, \alpha_j^{\uparrow} \right) + V_2^{eff} \left( |
  \x^f_i -  \x^f_j |; \alpha_i^{\uparrow},
  \alpha_j^{\downarrow} \right).
\end{equation}

\begin{figure}[htb]
  \begin{center}
    $\begin{array}{c}
      \multicolumn{1}{l}{\mbox{\bf (a)}} \\ [-0.53cm]
      \includegraphics[width=0.48\textwidth]{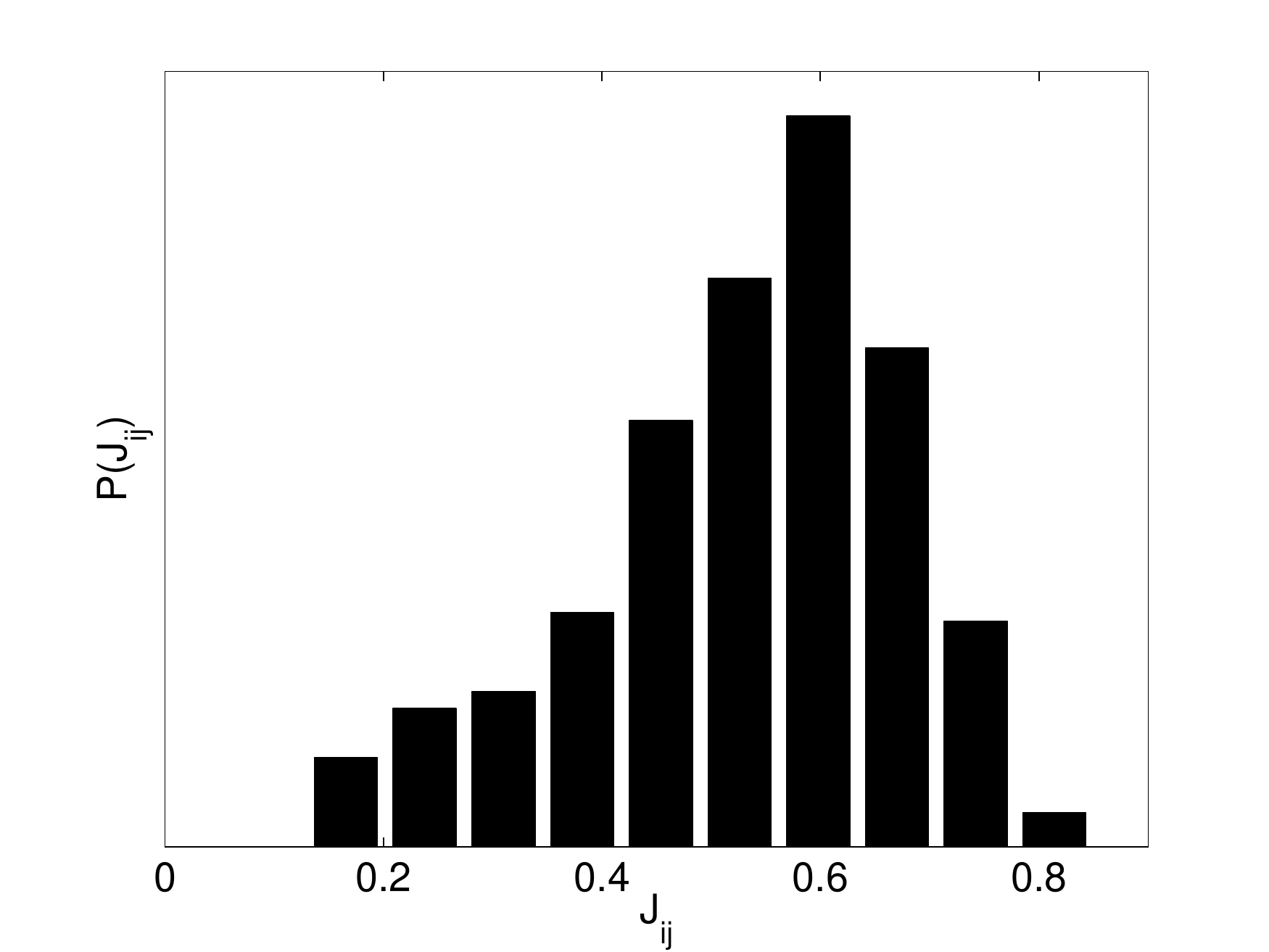} \\
      \multicolumn{1}{l}{\mbox{\bf (b)}} \\ [-0.53cm]
      \includegraphics[width=0.48\textwidth]{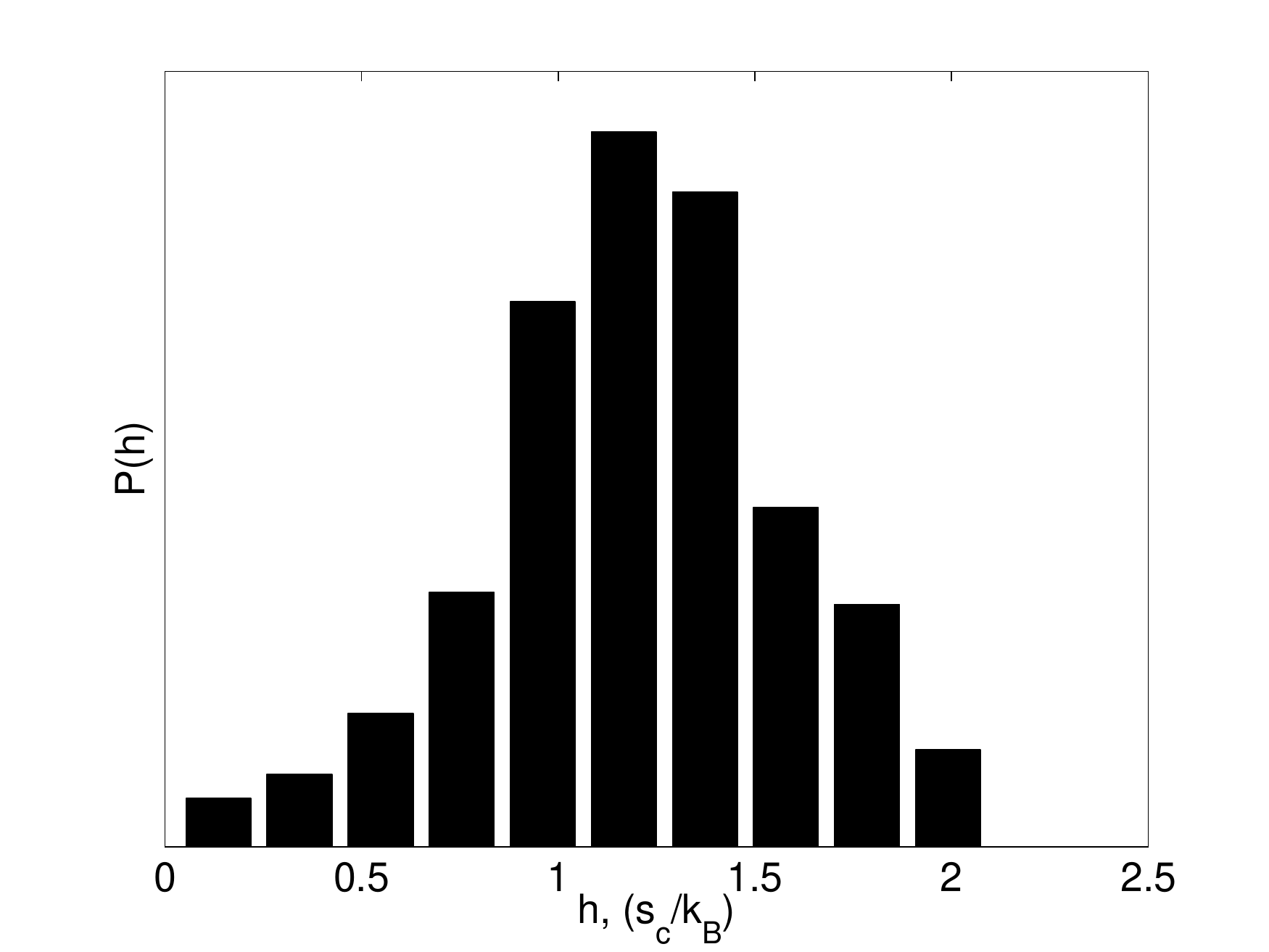} \\
    \end{array}$
  \end{center}
  \caption{The distributions of interactions and local fields of the magnet
  analogous to the simulated LJ two compound glass.  In this mapping $\bar{h}$
  is directly related to the configurational entropy, $\bar{h} = s_c / k_B$.
  The fields are shown at $\bar{h}=1.2$, close to the dynamical crossover
  temperature.}
  \label{fig:hijij}
\end{figure}

\section{Application to a simulated glass}
\label{sec:results}

\begin{figure*}[htb]
  \begin{center}
    \includegraphics[width=0.68\textwidth]{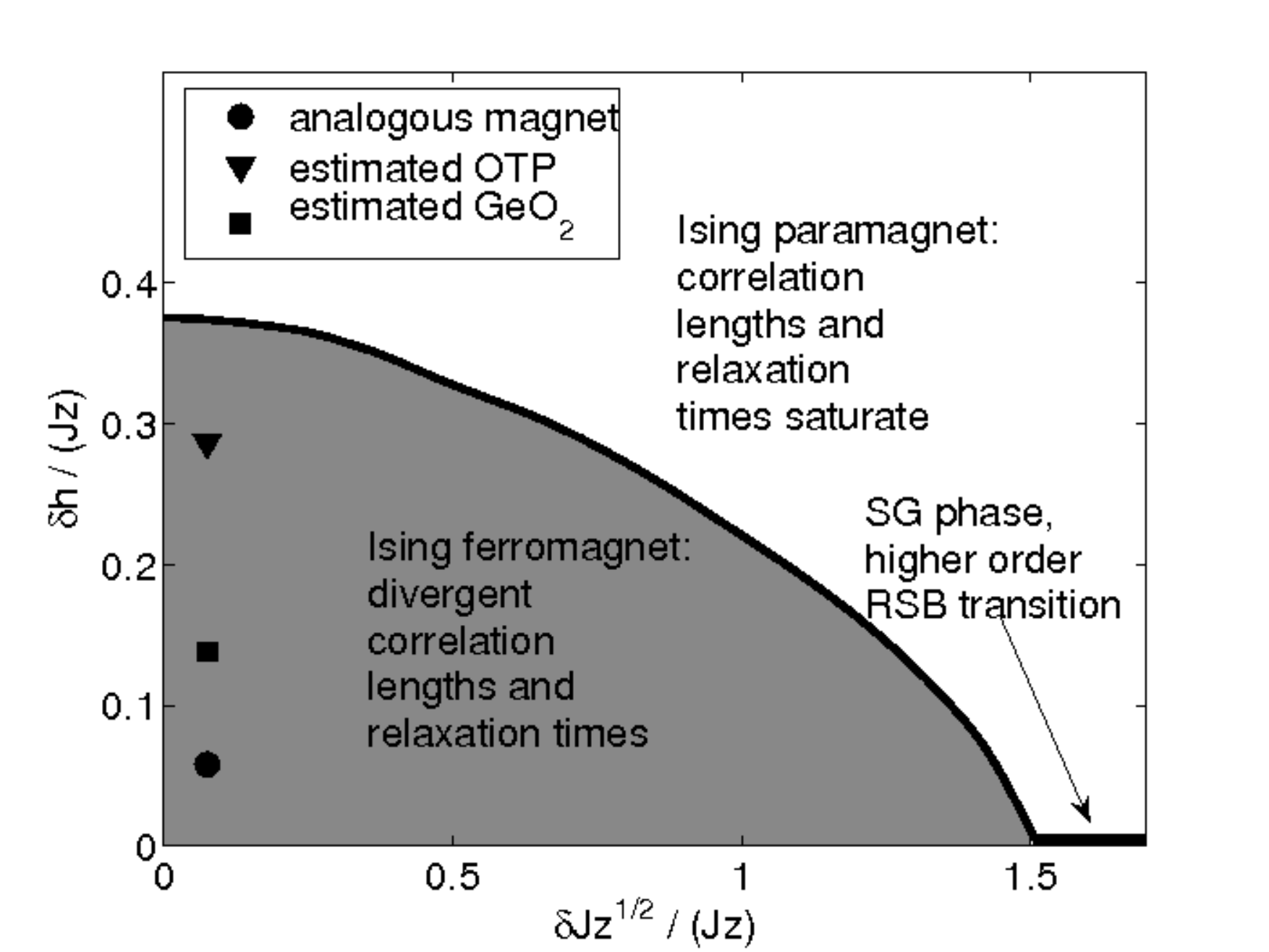}
  \end{center}
  \caption{Phase diagram of the Ising model with random bonds and fields
  adapted from reference \protect{\cite{migliorini.1998}}.  The parameters
  calculated for the magnet analogous to the LJ liquid (circular mark) indicate
  that the liquid would undergo a true phase transition at the ideal glass
  transition.  The triangular and square marks indicate estimates of where the
  glass forming liquids OTP and GeO$_2$ would fall on the phase diagram.}
  \label{fig:berker}
\end{figure*}

The mapping to the disordered Ising model should be carried out for each
fiducial equilibrium liquid structure.  We sample fiducial structures of the
Kob-Andersen 80-20 mixture of two types of Lennard-Jones (LJ) particles at
density $1.2$ (in LJ reduced units). The pairs have interaction
parameters\cite{kob.1994,hall.2008}, $\sigma_{AA} = 1.0$, $\sigma_{BB} = 0.8$,
$\sigma_{AB} = 0.88$, $\epsilon_{AA} = 1.0$, $\epsilon_{BB} = 1.5$,
$\epsilon_{AB} = 0.5$.  The fiducial structures were obtained by simulated
annealing runs to the temperature $T_{MD} = 0.45$. The equilibration time for
these simulations reaches a tenth of a microsecond when referenced to argon.

The parameters $\{ \alpha_i^{\uparrow} \}$ ($\{ q_i^{\uparrow} \}$) and $\{
\alpha_i^{\downarrow} \}$ ($\{ q_i^{\downarrow} \}$) are calculated for every
particle.  The mean RMS deviation determined from $\{\alpha_i\}$ is about
$0.12$ particle spacings, rather close to the Lindemann parameter expected for
periodic crystals $d_L \approx 0.1$ independent of the force law.  The RMS
actually observed during the MD run at $T_{MD} = 0.45$ is consistent with this
estimate $d_L = 0.113$.

The distributions of calculated interactions are shown in figure
\ref{fig:hijij}a as the free energy per neighbor, $J_i \equiv
\frac{1}{z_i} \sum_j J_{ij}$ where $z_i$ is the number of neighbors of particle
$i$.  The typical interaction $\bar{J} \equiv \frac{1}{N} \sum_i J_i$ is
directly related to $\sigma$, the mismatch free energy penalty in RFOT theory
for a particle at a flat interface between regions of high and low overlap.  In
the Ising mapping $\beta \sigma_I = n_{bb} \bar{J}$ where $n_{bb} = 3.2$ is the
typical number of bonds broken by the interface.  The direct calculation yields
$\bar{J} = 0.55$ giving $\beta \sigma_I = 1.77$ not very different from the
RFOT theory estimate usually used\cite{xia.2001} $\sigma_{RFOT} = \frac{3}{4}
k_BT \ln \frac{1}{d_L^2 \pi e} = 1.85 k_B T$.

In RFOT theory the configurational entropy parametrizes a liquid's descent into
the glassy regime. In harmony with many experimental
observations\cite{stevenson.2006,lubchenko.2004,novikov.2003}, the dynamic
crossover and the laboratory glass transition occur at universal critical
entropies of $s_c(T_c) = 1.12k_B$ and $s_c(T_g) = 0.82k_B$, respectively.

Because of the rapidly increasing equilibration time scales it is impossible
presently to obtain proper fiducial structures directly at very low
temperatures via molecular dynamics.  We can, however, treat the
configurational entropy, and therefore the average field, as variable in order
to extrapolate to find the magnetic system analogous to a liquid equilibrated
at a much lower temperature, eventually extrapolating all the way to the ideal
glass transition by taking $\bar{h} = \frac{1}{N} \sum_i h_i \to 0$.  The
presence of even a small average field is thought to destroy the phase
transitions of both spin glasses and the random field Ising magnets, but at
zero field a transition to a phase with long range correlations can still
occur.  Would a transition occur for the liquid analog when $\bar{h}=0$, i.e.
when the mean field configurational entropy vanishes?  We answer this by
appealing to an RG analysis of Migliorini and Berker\cite{migliorini.1998} for
the phase diagram for an Ising system in which both the fields and the
interactions fluctuate randomly and independently, encompassing both the RFIM
and the short range spin glasses.  Their model is on a cubic lattice.  We
present their phase diagram in figure \ref{fig:berker}  in terms of the mean
field theory based normalization where both the field fluctuations, $\delta h$,
and the fluctuations of the interaction strength, $\delta J z^{1/2}$, are
normalized by the total interaction energy per site $\bar{J} z$.  This
parametrization should eliminate trivial near neighbor lattice dependence.
This zero average field phase diagram is shown at a temperature $T=1$
coinciding with the established temperature of the analogous magnet.  The
results for the simulated Kob-Andersen liquid are indicated by the dot
suggesting the disorder in both the fields and the interactions is sufficiently
modest so that the system would undergo a phase transition to a state with
infinite correlation lengths and divergent relaxation times when the field
vanishes.  This extrapolation implies the Kob-Andersen LJ liquid should possess
a true phase transition to a state with one step replica symmetry breaking
(RSB), even though the re-normalized configurational entropy, which would
include small scale droplet excitations, strictly speaking remains finite.  For
such a broken replica symmetry state the free energy landscape has divergent
barriers between a finite number of collective free energy basins each one of
which still has finite configurational entropy due to local defects.  

According to Landau, the excess heat capacity\cite{landau.1969} yields $\delta
s_c = \sqrt{\Delta C_p k_B / N_{corr}}$, where $N_{corr}$ is the volume within
which the disorder is correlated.  The explicit mapping for the LJ system gives
entropy fluctuations $\delta h = \delta s_c / k_B= 0.38$, smaller than a
typical fragile glass forming liquid such as ortho-terphenyl (OTP).
Assuming a similar distribution of interactions but rescaling the field
fluctuations to that calculated for ortho-terphenyl yields the triangular
position on the RG phase diagram of figure \ref{fig:berker} while the square
mark indicates where the strong glass forming liquid GeO$_2$ would lie. Even
though OTP is rather fragile, we see it would still be expected to obey one
step RSB.

\begin{figure}[tpb]
  \begin{center}
    $\begin{array}{c}
      \multicolumn{1}{l}{\mbox{\bf (a)}} \\ [-0.53cm]
      \includegraphics[width=0.48\textwidth]{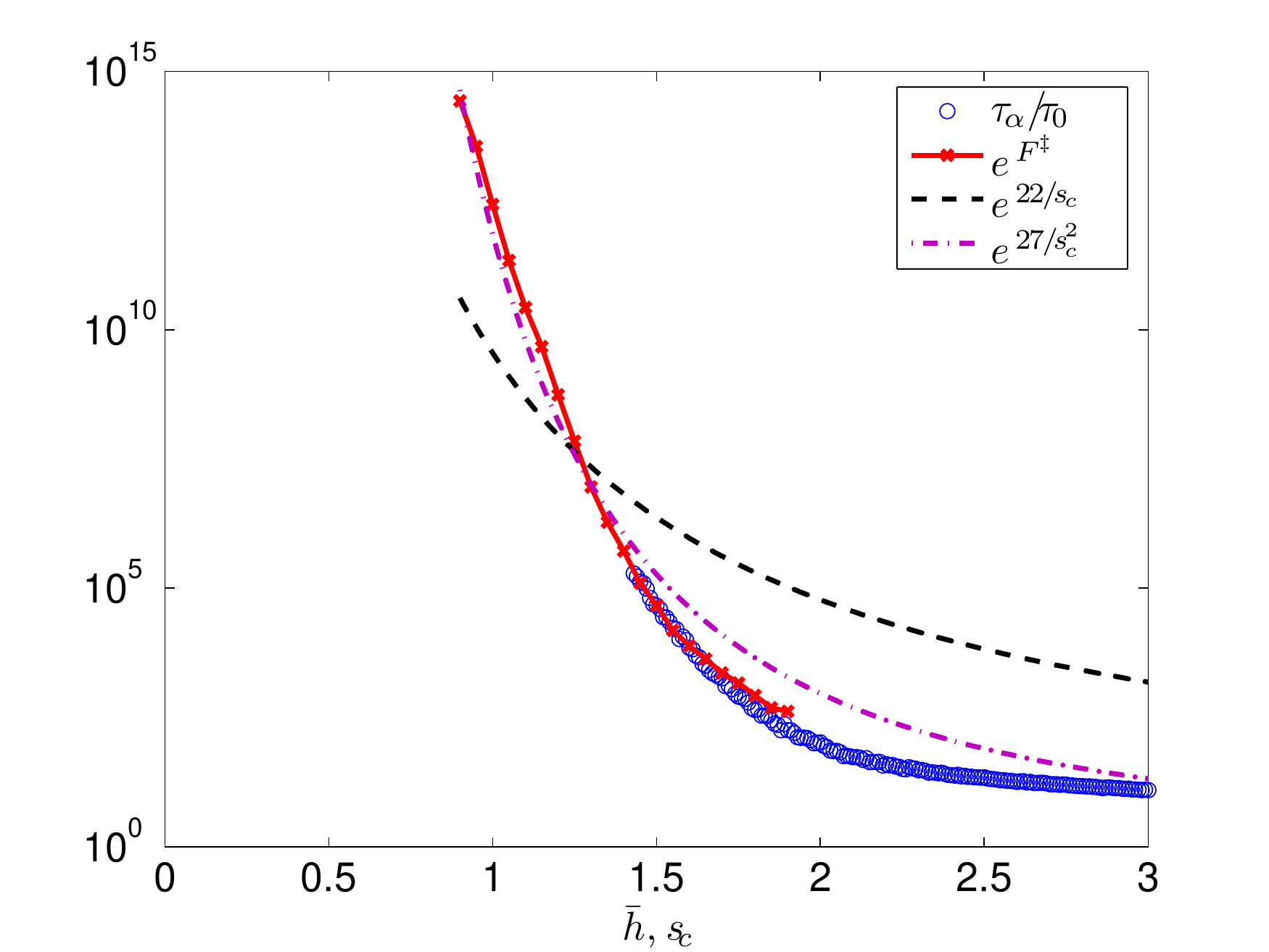} \\
      \multicolumn{1}{l}{\mbox{\bf (b)}} \\ [-0.53cm]
      \includegraphics[width=0.48\textwidth]{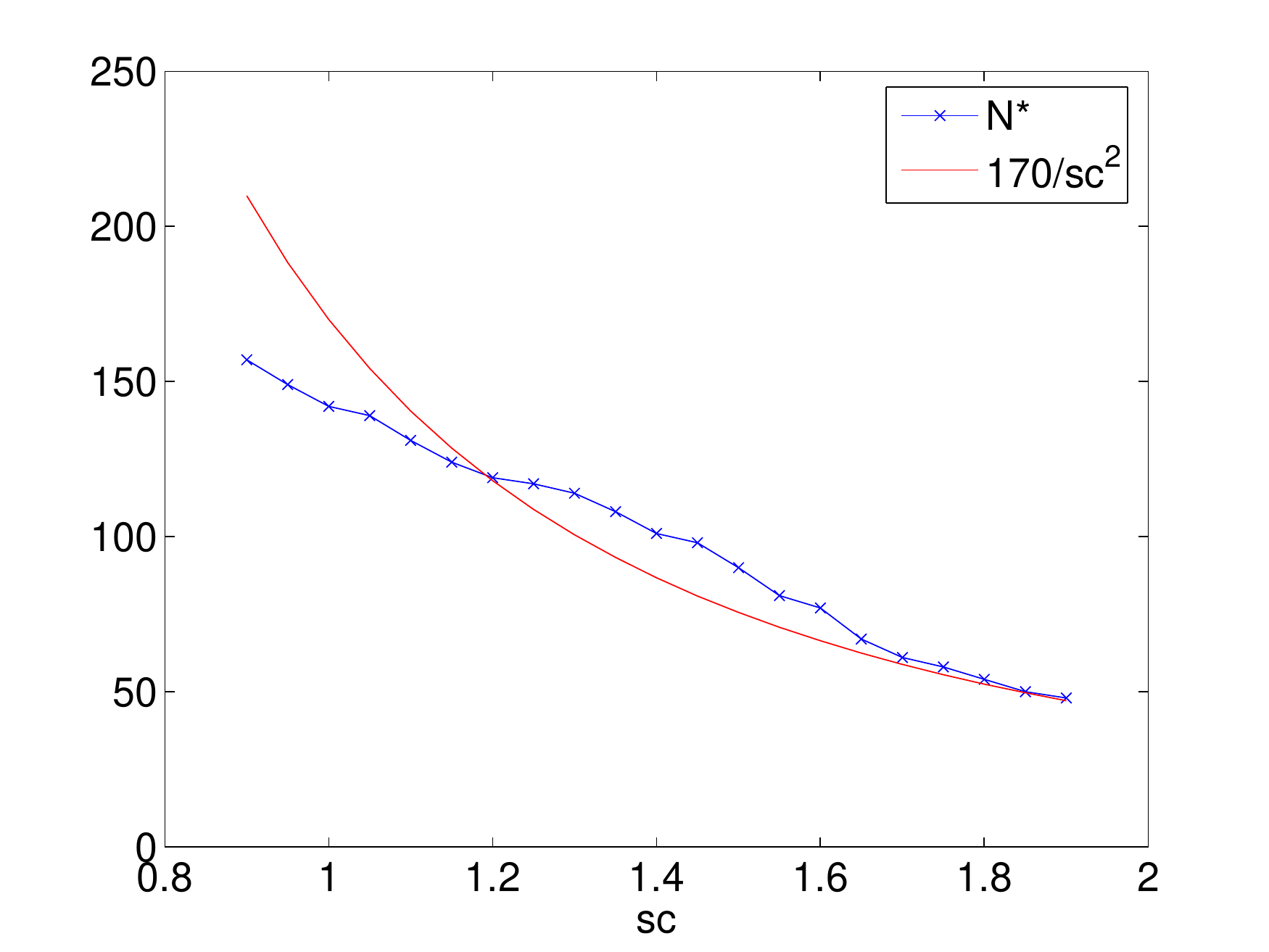} \\
    \end{array}$
  \end{center}
  \caption{(color online) (a) Relaxation times of the Ising model analogous to the LJ liquid
  (circles).  The solid line gives relaxation times calculated from free energy
  barriers.  The dashed lines show fits using relations derived in RFOT theory
  (see text).  (b) The minimum region size able to irreversibly reconfigure.}
  \label{fig:tau}
\end{figure}

Our extrapolations neglect any structural changes that occur in an actual fluid
upon cooling yet we can test how well the direct dynamics of the extrapolated
analog model correspond to droplet analysis. Escape from the metastable ($s=1$)
large overlap state corresponds with a large scale activated, structural
rearrangement of the liquid.  Directly simulated escape times are shown as
circles in  figure \ref{fig:tau}a.  The relaxation time grows rapidly at the
dynamical crossover temperature appearing to diverge as $s_c, \bar{h} \to 0$.
Below the dynamic crossover the growth in relaxation time is well fit by $\ln
\tau / \tau_0 \sim s_c^{-\psi}$.  The proportionality constant for the inverse
linear fit ($\psi = 1$) is $22k_B$, while droplet arguments in RFOT theory
predict a slightly larger value $\ln \tau / \tau_0 = 32k_B / s_c$.  $\psi = 2$,
corresponding to the unwetted result from RFOT theory\cite{xia.2000}, actually
gives a closer fit to the relaxation time curve.  This is consistent with what
we have already seen in figure \ref{fig:berker}, that the analog magnet
underestimates the disorder in the field moving the system away from the
critical line where wetting is dominant. 

The average overlap of the liquid frozen density fields maps onto the
magnetization in the analog magnet, $q=\frac{1}{N} \sum_i s_i$.  This
coordinate can be used to monitor escape from local minima.  We create free
energy profiles for this local collective reaction coordinate using the
weighted histogram analysis
method\cite{ferrenberg.1988,ferrenberg.1989,drozdov.2004} (WHAM).  The
resulting free energy profile, calculated with $\bar{h} = s_c /k_B = 1.1$, is
shown as a thick solid line in figure \ref{fig:Fqschematic}b.  The metastable
minimum at large overlap is separated from the global minimum at small overlap
by a free energy barrier that accounts for the relaxation time according to
$\tau = \tau_0 e^{\beta F^{\ddagger}}$.

The global overlap is not an ideal reaction coordinate for reconfiguration, as
it averages over reconfiguration events occurring at spatially distinct
regions.  By selecting a spherical region at random and only permitting motion
within that region, the overlap becomes a good reaction coordinate. This is the
magnetic analogy of the landscape ``library construction''\cite{lubchenko.2004}
and is similar to a technique recently used to calculate the surface tension
near a first order transition\cite{cammarota.2007}.  By varying the region
size, the minimum size to irreversibly escape a minimum and reconfigure the
liquid, $N^{*}$, can be determined.  Free energy profiles for several region
sizes around $N^{*}$ are shown in figure \ref{fig:Fqschematic}b.  The free
energy barriers computed for regions of size $N^{*}$, converted to relaxation
time, are shown in figure \ref{fig:tau}a.  Using WHAM and the library
construction allows descent much further into the glassy regime than is
possible via direct simulation.   The predicted minimum reconfiguration size is
shown in figure \ref{fig:tau}b. At high temperatures, $s_c > 1$, the growth of
region size with decreasing $s_c$ is consistent with $N^{*} \propto s_c^{-2}$
expected from RFOT theory, but at low temperatures the growth falls off as the
result of finite size effects since the cluster size approaches the simulated
system size itself.  The free energy barrier is also underestimated for the low
temperature range.

\begin{figure}[tpb]
  \begin{center}
    \includegraphics[width=0.48\textwidth]{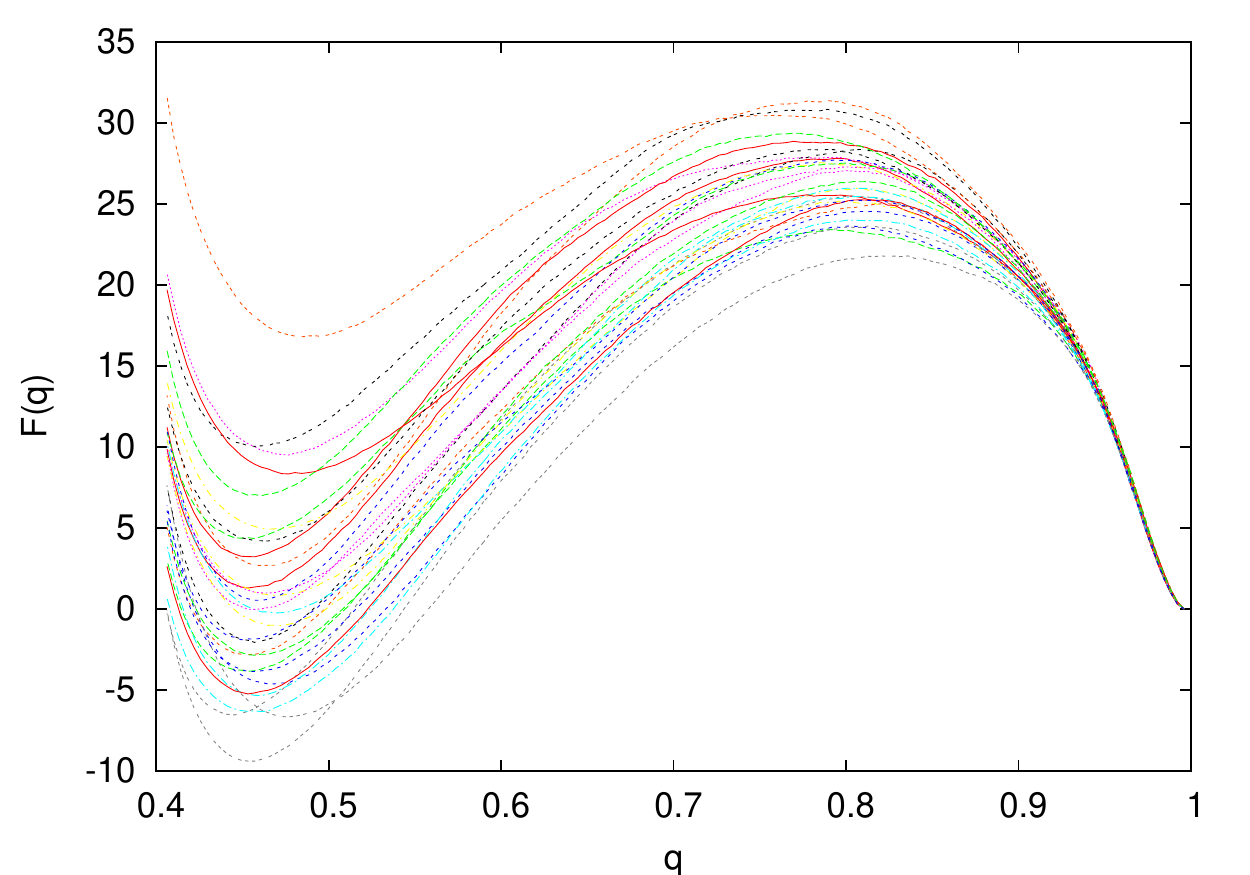}
  \end{center}
  \caption{(color online) Free energy profiles for different regions at $\bar{h}  = s_c /k_B =
  1.1$.  The distribution of free energy barriers gives rise to the stretched
  exponential relaxation behavior common to glassy systems.}
  \label{fig:heterogeneity}
\end{figure}

Not all cooperatively rearranging regions are created equal.  The resulting
dynamic heterogeneity of the liquid is seen in figure \ref{fig:heterogeneity}
showing a collection of free energy profiles for different regions at $\bar{h}
= s_c / k_B = 1.1$.  There is clearly a spread of relaxation times which can
give rise to the stretched exponential relaxation behavior $\phi (t) = e^{
-(t/\tau)^{ \beta_{KWW} }}$ common to glassy systems. If the relaxation is
entirely heterogeneous the stretching exponent, $\beta_{KWW}$, was shown in
reference \cite{xia.2001} to be related to the spread of free energy barriers,
$\delta F^{\ddagger}$, through the relation $\beta_{KWW} \approx \left( 1 +
(\delta F^{\ddagger} / k_BT )^{2} \right)^{-1/2}$.  Xia and Wolynes argued that
regions do not reconfigure completely independently, so free energy barriers
larger than the mean are lowered by facilitation effects of neighboring
regions.  Using the barrier distribution corrected in this way for facilitation
yields a non-exponentiality parameter nearly independent of temperature,
$\beta_{KWW} \approx 0.6$, a value characteristic for fragile liquids.

\section{Conclusion}

Using a mixed density functional/atomistic replica formalism, the dynamics of
the structural glass forming liquids can be mapped onto a general disordered
Ising model.  This mapping allows a computationally inexpensive route to low
temperature dynamics impossible currently by direct simulation.  Using this
information from the replica density functional one can guide molecular
dynamics simulations carried out at complete atomic detail to more easily reach
low temperature structural states. 

Our results suggest the Kob-Andersen liquid should demonstrate one step replica
symmetry breaking at a sufficiently low temperature even though its
configurational entropy including local droplet excitations, strictly speaking,
will not vanish.  The system is, in this sense, closer to the random field
Ising magnet than it is to the Edwards Anderson short range spin glass model.
The results from the simulation are consistent with droplet based predictions
using the existing random first order transition theory estimates.

\bigskip 

\begin{acknowledgments} 
  The authors thank David Sherrington for enjoyable discussions. Calculations
  were performed on computational facilities provided by LSU
  (http://www.hpc.lsu.edu) and the Louisiana Optical Network Initiative
  (http://www.loni.org).  Support from NSF grant CHE0317017 and NIH grant
  5R01GM44557 is gratefully acknowledged.
\end{acknowledgments}

\bibliography{/people/jake/latex_files/jakes_biblio}

\newpage
\end{document}